\documentclass[
    ,final            
  ]
  {aipproc}

\layoutstyle{6x9}
\usepackage{amsmath}

\begin{document}

\title{Possible evidence that pulsars are quark stars}

\classification{
97.60.Gb, 
21.65.Qr, 
97.60.Jd 
} \keywords      {Pulsars, Quark matter, Neutron stars}

\author{Renxin Xu}{
  address={Astronomy Department, School of Physics, Peking University,
           Beijing 100871, China}
}

\begin{abstract}
It is a pity that the real state of matter in pulsar-like stars is
still not determined confidently because of the uncertainty about
cold matter at supranuclear density, even 40 years after the
discovery of pulsar. Nuclear matter (related to neutron stars) is
one of the speculations for the inner constitution of pulsars even
from the Landau's time more than 70 years ago, but quark matter
(related to quark stars) is an alternative due to the fact of
asymptotic freedom of interaction between quarks as the standard
model of particle physics develops since 1960s. Therefore, one has
to focus on astrophysical observations in order to answer what the
nature of pulsars is. In this presentation, I would like to
summarize possible observational evidence/hints that pulsar-like
stars could be quark stars, and to address achievable clear evidence
for quark stars in the future experiments.
\end{abstract}

\maketitle

\noindent%
{\bf 1. Introduction: pulsars and cold quark matter}

It is a puerile desire to know the fundamental constituents of
matter and the interactions between them.
Elemental fermions (quarks and leptons) are supposed to be the
building blocks, between which fundamental interaction occurs via
exchanging gauge bosons, in the {\em standard model of particle
physics} that is one of most prominent achievements in the last
century. QCD (quantum chromo-dynamics) is believed to be the
underlying theory of the elementary strong interaction between
quarks, which is relatively poorly understood compared with others
(except for the planck-scale gravity).
QCD has been precisely tested in the high energy limit due to the
asymptotic freedom, while it becomes one of the daunting
challenges nowadays to understand QCD in the low energy regime
because of QCD's highly nonperturbative nature.
The aspects of nonperturbative QCD include the argument of color
confinement (would color states be really still single in
astrophysics?~\cite{cx06}) and possible phases of quark matter
(i.e., matter composed of quarks as building fermions).

It is a simple motivation idea to explore the Universe.
One of the attractive questions is about stars: their formation,
evolution, and death.
%
Pulsars, discovered 40 years ago, are the residues of massive main
sequent stars during supernovae, which were suggested to be normal
neutron stars (the most original one was speculated by Landau more
than 70 years ago) soon after the discovery, but were suspected to
be quark stars (QSs) as the quark model for hadrons develops since
1960s.
Pulsars are one kind of so-called compact stars, whose average
density is beyond the nuclear density, $\rho_{\rm nucl}$.
Thanks to advanced facilities, very different manifestations of
pulsar-like stars are observed recently, and much important
astrophysics depends certainly on the real nature of pulsars.

Both particle physicists and astrophysicists meet together when
trying to answer what the nature of matter at supranuclear density
is.
Neutron matter was proposed long before our recognizing the
structure of hadrons, and could certainly exist if neutron (and
protons, etc.) could be treated as point-like particles at the
energy scale of supranuclear density.
Quark matter is a direct consequence and a prediction of asymptotic
freedom, which could actually classified into two kinds: temperature
effect dominated (quark-gluon plasma, QGP) and density effect
dominated. The latter is relevant to pulsars (the focus of this
paper), the physics of which is illustrated in Fig.1 for a very
simple composition of protons and electrons.
\begin{figure}[h]
\includegraphics*[width=.9\textwidth]{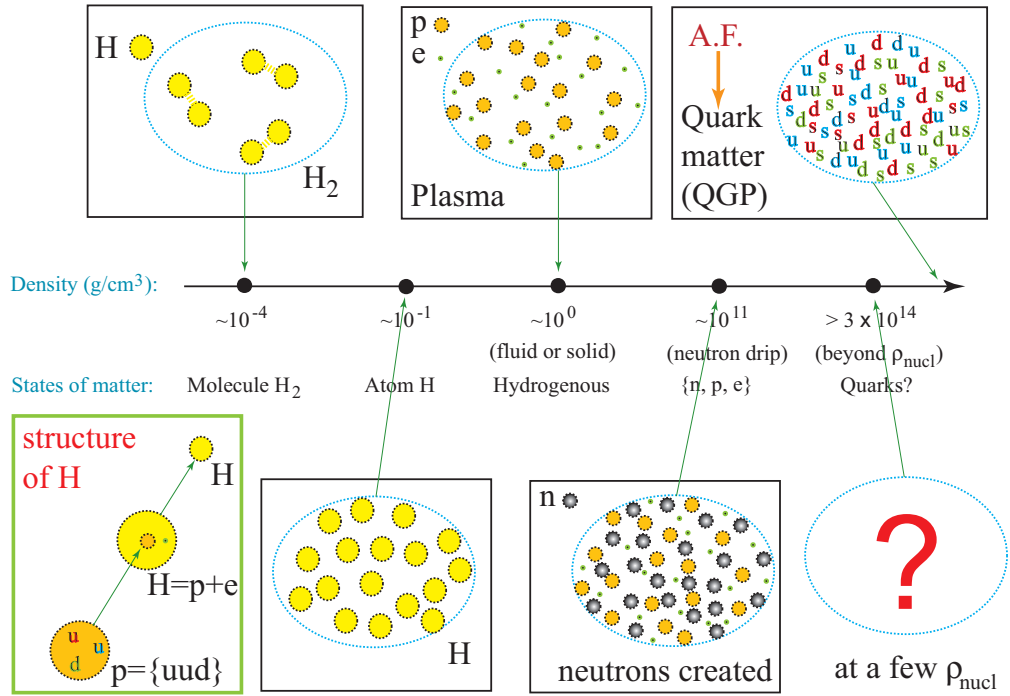}
\caption{Density effect dominated phases of matter composed simply
of electrons and protons. Temperature effect is negligible here.
It is worth noting that one could know the microphysics by
studying the states of matter at different densities. ``A.F.'':
asymptotic freedom, ``$\rho_{\rm nucl}$'': the nuclear density.}
\end{figure}
It is evident that, as the density increases, various states of
matter are identified. This is certainly an effective way to
reveal the fundamental constitutions and their interactions as
well.

The state of cold matter at a few $\rho_{\rm nucl}$ is still an
unsolved problem in nonperturbative QCD. Is it of neutron matter
or quark matter? What about cold quark matter?
Two kinds of efforts are made to date for the latter question.
($a$) Phenomenological models: to focus on pulsar-like stars as
astro-laboratories.
Based on different manifestations of pulsars, a solid state for
cold quark matter was conjectured by Xu~\cite{xu03}, who also
proposed more realistically that quark clusters could be essential
for the state called as {\em normal-solid}.
($b$) Effective models: to construct theoretical frames
qualitatively equipollent to QCD for specific problems.
BCS-type quark pairs may form at a Fermi surface of cold quark
matter, and the shear moduli of the rigid crystalline color
superconducting quark matter ({\em super-solid state}) could be 20
to 1000 times larger than those of neutron star
crusts~\cite{mrs07}.
It could be interesting to observationally distinguish between and
search evidence for possible normal-solid and super-solid states
although the latter might be more robust than the former from a
theoretical point of view.

Possible evidence for QSs will be summarized. My related previous
reviews~\cite{xu03a,xu03b,xu06,xu07} could be readable for
historical notes, backgrounds, and a few details.

\vspace{2mm}
\noindent%
{\bf 2. Possible evidence for quark stars}

QSs are conventionally thought as a special kind of neutron
stars~\cite{weber05}, but it is worth to distinguish them from other
normal neutron stars because of no free neutron and a very exotic
state in their interiors.
Although QSs seem to be `easily' ruled out from time to time in
the literatures (similar to the case of one's refraining from
smoking), we would like to address possible evidence for them
since this is still an unsolved physical and astrophysical problem
mixed with a variety of research subjects.

{\em Quark stars: to be bare?}
Radio pulsars were alternatively suggested to be crusted strange
stars~\cite{afo86} until Xu \& Qiao~\cite{xq98} argued the
magnetospheric activity of {\em bare} strange stars (BSSs) and
addressed three advantages for BSSs as the nature of puslars:
binding energy, spectral feature, and sucessful supernavae.
The RS-type vacuum gap model~\cite{rs75}, with an ``user
friendly'' nature, is popular and successful in explaining the
radiative behaviors of radio pulsars, which can only work in
strick conditions: strong magnetic field and low temperature on
surface of pulsars with $\Omega\cdot{\bf B}<0$ (see,
e.g.,~\cite{ml07}). This binding energy could be completely solved
for any $\Omega\cdot{\bf B}$ if radio pulsars are
BSSs~\cite{xq98,xqz99}.
Drifting subpulses~\cite{dr99} and microstructures could be strong
evidence for RS-type sparking on polar caps, and further more, the
bi-drifting phenomena~\cite{qiao04,Bha07} could only be understood
in BSS models.

Besides RS-type sparking, the bare quark surface could also help
to explain a few other observations.
Only a layer of degenerated electrons in strong magnetic fields on
bare quark surface, which can naturally reproduce non-atomic
spectra~\cite{xu02} though atomic features were predicted in
normal neutron star models long before observations. The
absorption lines of several X-ray sources (e.g., 1E1207 and
SGR1806) could originate from transition between Landau levels of
electrons~\cite{xwq03}.
Additionally, the quark surface may help to alleviate the current
difficulties in reproducing two kinds of astronomical bursts which
are challenging today's astrophysicists to find realistic explosive
mechanisms. Because of chromatic confinement (the photon luminosity
of a quark surface is then not limited by the Eddington limit), BSSs
could create a lepton-dominated fireball~\cite{xu05,ph05} which
could push the overlying matter away through photon-electron
scattering with energy as much as $\sim 10^{51}$ erg for successful
supernovae~\cite{cyx07}. Asymmetric explosion in such a way may
naturally result in long-soft $\gamma$-ray bursts and in kicks on
QSs~\cite{cen98,Huang03,cui07}.

{\em Mass-radius relation: low-mass QSs?}
The striking difference between the mass-radius
relations~\cite{lp04} of normal neutron stars and (bare) QSs is
thought to be useful for identifying QSs (e.g. ~\cite{Li99}), and
yet it is worth paying attention to {\em low}-mass QSs~\cite{xu05}
since QSs and normal neutron stars with similar maximum mass can
hardly be distinguished observationally.
Specially, BSSs can be of very low mass (e.g., $\sim
10^{-4}M_\odot$, and radius $<1$ km), while normal neutron stars
cannot (the minimum mass $\sim 0.1M_\odot$, and radius $\sim 160$
km). In principle, the low limit of BSS's mass could almost be
zero because of self-confinement (quark nuggets, quark planets,
and QSs).
Solar-mass and low-mass QSs may form in different channels:
core-collapse explosion for the former and AIC (accretion-induced
collapse) of white dwarfs for the latter. The latter could also be
possibly the residue of cosmic QCD phase separation in the early
Universe.
Note that the astrophysical appearance of low-mass BSSs in both
rotation- and accretion-powered phases is quite different to the
standard scenarios~\cite{xu05}.

Actually, there may be some observational hints of low-mass
pulsar-like stars, which include the spin and polarization
behaviors (PSR 1937+21)~\cite{xxw01}, the peculiar timing behavior
(1E1207)~\cite{xu05}, no-detection of gravitational waves from
radio pulsars~\cite{xu06gw}, and small polar cap area (PSR
B0943+10)~\cite{ycx06}.
The detected small thermal area~\cite{pavlov04} ({\em if} being
global) of central compact objects may reflect their low masses too.

{\em Conjecture: solid quark matter?}
Based on a variety of observational features (Planck-like thermal
spectra, precessional movtion, glitches), a solid state of cold
quark matter is conjectured~\cite{xu03}, that could not be ruled
out by first principles of QCD. Quark clustering is necessary if
the solid is one kind of normal solid at low temperature.
%
This idea could explain naturally the discrepancy between precession
and glitch of radio pulsars.

As a solid QS evolves (initially cooling and solidification,
spinning down, accreting matter), strain develops and a star-quake
would occur if the stress increases to a critical value.
Quakes of solid QSs may result in pulsar glitches~\cite{z04} and
bursts (even super-flares) of anomalous X-ray pulsars/soft gamma-ray
repeaters~\cite{xty06}.
Actually there are two kinds of stress force inside solid stars:
bulk-invariable and bulk-variable forces, both of which could result
in decreases of moment of inertia, and thus in pulsar glitches.
The total stellar volume may keep almost constant if the former
dominates during quake (see Fig.2 for a demonstration), but not if
the latter dominates.
\begin{figure}[h]
\includegraphics*[width=.9\textwidth]{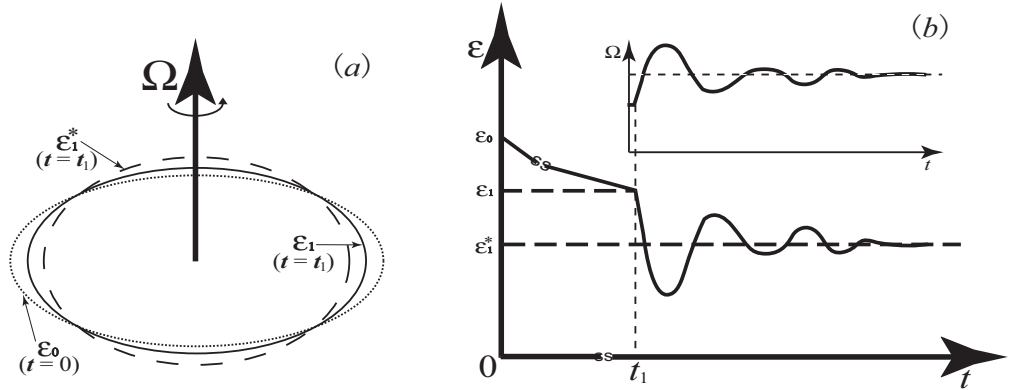}
\caption{An illustration of bulk-invariable force induced quake of
a solid quark star. The Maclaurin figure determines the
eccentricities of $\varepsilon_0$ and $\varepsilon_1^*$. A
star-quake with real eccentricity $\varepsilon_1$ occurs at time
$t_1$.}
\end{figure}
Small or slow glitches could be relevant to bulk-invariable force,
and numerical calculations show that the observed slow glitches
could be reproduced for a certain parameter space~\cite{px07}.

\vspace{2mm}
\noindent%
{\bf 3. Evidence in the future?}

{\em Searching for sub-millisecond pulsars.}
Among the likely ways to identify QSs, discovering sub-millisecond
pulsars would model-independently be a clear evidence for
QSs~\cite{xu06subms} because normal neutron stars are bound by
gravitation while BSSs are chromatically confined.
The smallest spin period is $\sim 0.5M_1^{-1/2}R_6^{3/2}$ ms for
neutron stars with maximum mass $M_1 M_\odot$ and smallest radius
$R_6 10^6$ cm, by which the spin of QSs are not limited.
The low limit is higher if possible instabilities (e.g., the
$r$-mode one) are considered.
Recent observation may hint a discovery of QS~\cite{ZhangCM07}, in
fact.

There could be several ways to form sub-millisecond radio pulsars.
$a$, A white dwarf is spun up to a high state of angular momentum
in an accreting binary, and would collapsed to be a
sum-millisecond pulsar after AIC.
$b$, Merging quark nuggets soon after cosmic QCD phase transition
may result in low-mass QSs with high spins.
$c$, Orbital angular momentum is transferred into spin one of
compact low-mass QSs during accretion phase of binaries.
$d$, The mass of a BSS with mass of $\sim 10^{-4}M_\odot$ may
increase to $\sim 10^{-3}M_\odot$ after $\sim 10^6$-year's
accretion with Eddington rate.
These mean that BSSs with a few kilometers in radius and with
sum-millisecond periods are possible after significant accretion
in binaries.
Should a sub-millisecond radio pulsar be a QS with very low mass,
its emissivity would be weak because of small both moment of
inertia and magnetic moment. We need large telescope (e.g.,
$FAST$) to find such radio pulsars in order to cool down the long
heated debate on the nature of pulsars.

{\em Others.}
It would be direct and important to measure certainly a small
bolometric radius (e.g., $<5$ km) of QS in X-ray (e.g.,
Constellation-X) and UV bands (to detect emission component at low
energy).
Is there any hint of QSs in both accretion and no-accretion
binaries? It is worth searching observational QS-features in
binaries with companies of white dwarfs, giants or super-giants
(e.g., the symbiotic X-ray binaries), main-sequent stars, etc.
Besides, QSs and normal neutron stars could be differentiated by
their radiative features of gravitational waves, which has been
studied diversely~\cite{Owen05,Horvath05,Lin07,ligo07}.
It is worth noting no-detection may hint a low mass of QS since the
gravitational wave behaviors should be mass-dependent~\cite{xu06gw}.
Additionally, detecting strangelets as cosmic rays (e.g., in the
future AMS02 experiment) could also be in-direct evidence for quark
stars. A support~\cite{xu07} for solid quark matter could be through
identifying quark clusters in strongly coupled QGP created in
relativistic colliders (e.g., RHIC).

\vspace{2mm}
\noindent%
{\bf 4. Conclusions}

{\em Low-mass solid bare} strange quark stars could be the nature
of pulsars, that are {\em not} ruled out and would probably be
discovered in the future.
I focused on the work of my group and feel sorry for neglecting
many interesting references due to the page limit.

\vspace{2mm}
\noindent%
{\bf Acknowledgments}:
I acknowledge the contributions by my colleagues at the pulsar
group. The work is supported by NSFC (10573002, 10778611), the Key
Grant Project of Chinese Ministry of Education (305001), and by
LCWR (LHXZ200602).


\begin{thebibliography}{}

\bibitem{ligo07}
B. Abbott, et al., \emph{All-sky search for periodic gravitational
waves in LIGO S4 data}, preprint (arXiv:0708.3818), 2007.

\bibitem{afo86}
C. Alcock, E. Farhi, A. Olinto, \emph{Strange stars}, ApJ, 310,
261-272, 1986.

\bibitem{Bha07}
B. Bhattacharyya, Y. Gupta, J. Gil, M. Sendyk, \emph{Discovery of
a remarkable subpulse drift pattern in PSR B0818-41}, MNRAS, 377,
L10-L14, 2007.

\bibitem{cen98}
R. Y. Cen, \emph{Supernovae, Pulsars, and Gamma-Ray Bursts: A
Unified Picture}, ApJ, 507, L131-L134, 1998.

\bibitem{cyx07}
A. B. Chen, T. H. Yu, R. X. Xu, \emph{The Birth of Quark Stars:
Photon-driven Supernovae?}, ApJ, in press (astro-ph/0605285),
2007.

\bibitem{cui07}
X. H. Cui, H. G. Wang, R. X. Xu, G. J. Qiao, \emph{Pulsar kicks
and $\gamma$-ray burst}, A\&A, 472, 1-3, 2007.

\bibitem{dr99}
A. Deshpande, J. Rankin, \emph{Pulsar magnetospheric emission
mapping: Images and implications of polar cap weather}, ApJ, 524,
1008-1013, 1999.

\bibitem{Huang03}
Y. F. Huang, Z. G Dai, T. Lu, K. S. Cheng, X. F. Wu,
\emph{Gamma-Ray Bursts from Neutron Star Kicks}, ApJ, 594,
919-923, 2003.

\bibitem{lp04}
J. M. Lattimer, M. Prakash, \emph{The Physics of Neutron Stars},
Science, 304, 536-542, 2004.

\bibitem{Li99}
X. D. Li, I. Bombaci, M. Dey, J. Dey, E. P. J. van den Heuvel,
\emph{Is SAX J1808.4-3658 a Strange Star?}, 83, 3776-3779, 1999.

\bibitem{mrs07}
M. Mannarelli, K. Rajagopal, R. Sharma, \emph{The rigidity of
crystalline color superconducting quark matter}, preprint
(hep-ph/0702021).

\bibitem{Owen05}
B. J. Owen, \emph{Maximum Elastic Deformations of Compact Stars
with Exotic Equations of State}, Phys. Rev. Lett. 95, 211101,
2005.

\bibitem{ph05}
B. Paczy\'nski, P. Haensel, \emph{Gamma-ray bursts from quark
stars}, MNRAS, 362, L4-L7, 2005.

\bibitem{pavlov04}
G. G. Pavlov, D. Sanwal, M. A. Teter, \emph{Central Compact
Objects in Supernova Remnants}, in: Young Neutron Stars and Their
Environments, IAU Symposium no. 218, pp.239, 2004.

\bibitem{px07}
C. Peng, R. X. Xu, \emph{Pulsar slow glitches in a solid quark
star model} (arXiv:0708.2482), 2007.

\bibitem{qiao04}
G. J. Qiao, K. J. Lee, B. Zhang, R. X. Xu, H. G. Wang, \emph{A
Model for the Challenging "Bi-drifting" Phenomenon in PSR
J0815+09}, ApJ, 616, L127-L130, 2004.

\bibitem{cx06}
C. X. Qiu and R. X. Xu, \emph{Color-charged Quark Matter in
Astrophysics?}, Chin. Phys. Lett., 23, 3205-3207, 2006
(astro-ph/0608272).

\bibitem{Horvath05}
J. E. Horvath, \emph{Energetics of the Superflare from SGR1806-20
and a Possible Associated Gravitational Wave Burst}, Modern
Physics Letters A, 20, 2799-2804, 2005 (astro-ph/0508223).

\bibitem{Lin07}
Lap-Ming Lin, \emph{Constraining Crystalline Color Superconducting
Quark Matter with Gravitational-Wave Data}, Phys. Rev. D, in press
(arXiv:0708.2965), 2007.

\bibitem{ml07}
Z. Medin, D. Lai, \emph{Condensed Surfaces of Magnetic Neutron
Stars, Thermal Surface Emission, and Particle Acceleration Above
Pulsar Polar Caps}, MNRAS, submitted (arXiv:0708.3863).

\bibitem{rs75}
M. A. Ruderman and P. G. Sutherland, \emph{Theory of pulsars -
Polar caps, sparks, and coherent microwave radiation}, ApJ, 196,
51-72, 1975.

\bibitem{weber05}
F. Weber, \emph{Strange quark matter and compact stars}, Prog.
Part. Nucl. Phys., 54, 193-288, 2005.

\bibitem{xu02}
R. X. Xu, \emph{A Thermal featureless spectrum: Evidence for bare
strange stars?}, ApJ, 570, L65-L68, 2002.

\bibitem{xu03}
R. X. Xu, \emph{Solid quark stars?}, ApJ, 596, L59-L62, 2003.

\bibitem{xu03a}
R. X. Xu,, \emph{Strange Quark Stars - A Review}, in: High Energy
Processes, Phenomena in Astrophysics, Proceedings of IAU Symposium
No. 214, eds. X. D.Li, Z. R. Wang \& V. Trimble, 191-198, 2003.

\bibitem{xu03b}
R. X. Xu,, \emph{Bare strange stars: formation and emission}, in:
Stellar astrophysics --- a tribute to Helmut A. Abt, eds. K. S.
Cheng,  K. C. Leung \& T. P. Li , Kluwer Academic Publishers,
73-81, 2003.

\bibitem{xu05}
R. X. Xu,, \emph{1E 1207.4-5209: a low-mass bare strange star?},
MNRAS, 356, 359-370, 2005.

\bibitem{xu06}
R. X. Xu,, \emph{Pulsars and Quark stars}, Chin. J. A\&A Suppl.,
2, 279-286, 2006 (astro-ph/0512519).

\bibitem{xu06gw}
R. X. Xu,, \emph{To probe into pulsar's interior through
gravitational waves}, Astropart. Phys., 25, 212-219, 2006
(astro-ph/0511612).

\bibitem{xu06subms}
R. X. Xu,, \emph{Searching for sub-millisecond pulsars: a
theoretical view}, in: Gravitation and Astrophysics: on the
occasion of the 90th year of General Relativity, eds. J. M.
Nester, C. M. Chen, \& J. P. Hsu, World Scientific, p.159-167,
2007 (astro-ph/0605028).

\bibitem{xu07}
R. X. Xu,, \emph{AXPs/SGRs: Magnetars or Quarkstars?}, Advances in
Space Research, in press (astro-ph/0611608), 2007.

\bibitem{xq98}
R. X. Xu, G. J. Qiao, \emph{``Bare'' strange stars might be not
bare}, Chin. Phys. Lett., 15, 934-936, 1998 (astro-ph/9811197).

\bibitem{xqz99}
R. X. Xu, G. J. Qiao, B. Zhang, \emph{PSR 0943+10: a bare strange
star?} ApJ, 522, L109-L112, 1999.

\bibitem{xty06}
R. X. Xu, D. J. Tao, Y. Yang, \emph{The superflares of soft gamma
ray repeaters: giant quakes in solid quark stars?}, MNRAS, 373,
L85-L89, 2006.

\bibitem{xwq03}
R. X. Xu, H. G. Wang, G. J. Qiao, \emph{A note on the discovery of
absorption features in 1E 1207.4-5209}, Chin. Phys. Lett., 20,
314-316, 2003 (astro-ph/0207079).

\bibitem{xxw01}
R. X. Xu, X. B. Xu, X. J. Wu, \emph{The Fastest Rotating Pulsar: A
Strange Star?}, Chin. Phys. Lett., 18, 837-840, 2001
(astro-ph/0101013).

\bibitem{ycx06}
Y. L. Yue, X. H. Cui, R. X. Xu, \emph{Is PSR B0943+10 a low-mass
quark star?}, ApJ, 649, L95-L98, 2006.

\bibitem{ZhangCM07}
C. M. Zhang, H. X. Yin, Y. H. Zhao, Y. C. Wei, X. D. Li,
\emph{Does Sub-millisecond Pulsar XTE J1739-285 Contain a Low
Magnetic Neutron Star or Quark Star?}, PASP, accepted
(arXiv:0708.3566), 2007.

\bibitem{z04}
A. Z. Zhou, R. X. Xu, X. J. Wu, N. Wang, \emph{Quakes in solid
quark stars}, Astropart. Phys., 22, 73-79, 2004
(astro-ph/0404554).

\end{thebibliography}
\end{document}